\documentclass[aps,superscriptaddress]{revtex4}
\usepackage{amsmath}
\usepackage{graphicx}

\begin{document}

\title{Adiabatic-Nonadiabatic Transition in the Diffusive Hamiltonian
Dynamics of a Classical Holstein Polaron}
\author{Alex A. Silvius}
\affiliation{Department of Physics, University of Missouri-Rolla, MO 65409}
\author{Paul E. Parris}
\affiliation{Department of Physics, University of Missouri-Rolla, MO 65409}
\affiliation{Laboratoire P.Painlev\'{e} and UFR de Math\'{e}matiques, Universit\'{e}
des Sciences et Technologies de Lille, 59655 Villeneuve d'Ascq, FRANCE}
\author{Stephan De Bi\`{e}vre}
\affiliation{Laboratoire P.Painlev\'{e} and UFR de Math\'{e}matiques, Universit\'{e}
des Sciences et Technologies de Lille, 59655 Villeneuve d'Ascq, FRANCE}

\begin{abstract}
We study the Hamiltonian dynamics of a free particle injected onto a chain
containing a periodic array of harmonic oscillators in thermal equilibrium.
The particle interacts locally with each oscillator, with an interaction
that is linear in the oscillator coordinate and independent of the
particle's position when it is within a finite interaction range. At long
times the particle exhibits diffusive motion, with an ensemble averaged
mean-squared displacement that is linear in time. The diffusion constant at
high temperatures follows a power law $D\sim T^{5/2}$ for all parameter
values studied. At low temperatures particle motion changes to a hopping
process in which the particle is bound for considerable periods of time to a
single oscillator before it is able to escape and explore the rest of the
chain. A different power law, $D\sim T^{3/4},$ emerges in this limit. A
thermal distribution of particles exhibits thermally activated diffusion at
low temperatures as a result of classically self-trapped polaronic states.
\end{abstract}

\date{\today }
\maketitle

\section{Introduction}

\label{intro}

The theoretical problem of an essentially free particle interacting locally
with vibrational modes of the medium through which it moves is relevant to a
large class of physical systems, and arises in a variety of contexts. It is
pertinent to fundamental studies aimed at understanding the emergence of
dissipation in Hamiltonian systems \cite{bdb,dissipation}, and to the
outstanding theoretical problem of deriving macroscopic transport laws from
the underlying microscopic dynamics \cite{cels,PhysicaD}. It appears,
perhaps most extensively, in the area of solid state physics \cite%
{ziman,feynman,Holstein,electronphonon}, where a great deal of theoretical
work has focused on the nature of electron-phonon interactions, and the rich
variety of behavior that occurs: from the simple scattering of particles by
phonons \cite{ziman}, to the emergence of polarons in which the itinerant
particle is clothed by a local distortion of the medium \cite%
{feynman,Holstein,electronphonon}, and even to self-trapped polaron states,
in which the particle is bound to the potential well created by such a
localized vibrational distortion. Such theoretical issues have also had a
rich interplay with experiment, with experimental observations suggesting
theoretically predicted band-to-hopping and adiabatic-nonadiabatic polaron
transitions of injected charges carriers in organic molecular solids \cite%
{scheinborsenberger}.

\begin{figure}[t]
\center \includegraphics[width=6in]{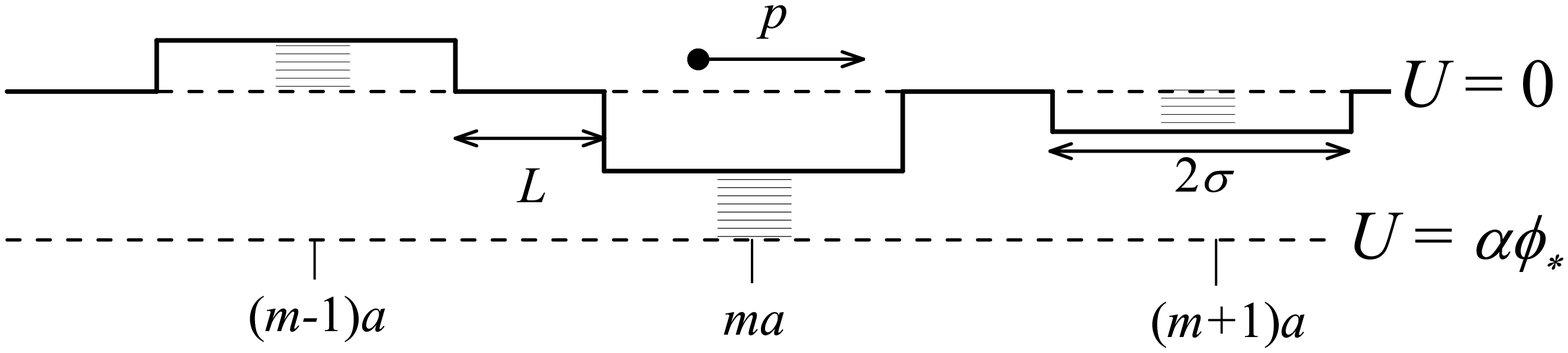}
\caption{Schematic illustration of the dynamical model. A particle moves
along a line in which are periodically embedded oscillators of frequency $%
\protect\omega$ which oscillate transverse to the particle motion. There is
no interaction when the particle is in one of the free regions of length $L$
between adjacent oscillators. The particle exerts a constant force on the
oscillator at $q_{m} = ma$ when it is within a distance $\protect\sigma$ of
that oscillator. As a result, the oscillator in that cell moves about a
shifted equilibrium position $\protect\phi_{\ast}=-\protect\alpha/\protect%
\omega^{2}.$ The thick line shows a snapshot of the instantaneous potential $%
U(q)$ experienced by the particle at one instant during the evolution of the
system. The shaded regions connect the potential in each interacting subcell
to the value about which it oscillates.}
\label{FIG1}
\end{figure}

Although much of the theoretical work on polaronic effects has appropriately
focused on quantum mechanical calculations of polaron hopping rates,
diffusion constants, and mobilities, there have been some workers who have
questioned the degree to which polaronic phenomena depend intrinsically on
their quantum mechanical underpinnings. There is an extensive literature,
e.g., dealing with the validity and usefulness of various semi-classical
approximations that have been used to study many aspects of polaron dynamics 
\cite{SemiClassical}. Considerably less work has focused on \emph{completely}
classical versions of the problem and the degree to which classical models
might capture essential transport features found in approximate or
perturbative quantum mechanical calculations \cite{oscillators}. Moreover,
since the exact quantum mechanical dynamics of the interacting
particle/many-oscillator system remains computationally complex due to the
enormous size of the many-body Hilbert space, it is difficult to
computationally test perturbative or semi-classical calculations of such
fundamental transport quantities as the polaron diffusion constant. It might
be hoped, therefore, that numerically exact calculations of the full
many-body dynamics of classical versions of the problem could provide
information that would be complementary to analytical studies and
semiclassical approximations of the full quantum evolution.

Motivated by these considerations, we study here the closed Hamiltonian
dynamics of a simple system that corresponds to a classical version of the
well-studied Holstein molecular crystal model \cite{Holstein}. The model, in
which a free mobile particle experiences a linear, local interaction with
vibrational modes of fixed frequency $\omega $ arranged at regular intervals 
$a$ along the line on which it moves, can also be viewed as a
one-dimensional periodic inelastic Lorentz gas.

Similar to what has been observed in quantum mechanical treatments of the
polaron problem we find that there are in this model two different types of
states of the combined system: self-trapped and itinerant. As their name
suggests, self-trapped states are immobile. Itinerant states, by contrast,
undergo motion that is macroscopically diffusive. We show for our classical
model that the diffusion constant for itinerant states has a power law
dependence on the temperature $T$, with two different regimes that are
similar to the adiabatic and non-adiabatic limits discussed in the polaron
literature. At high temperatures the diffusion constant of the model
increases with temperature as $T^{5/2}$ and the transport mechanism is
similar to what one thinks of as band transport in a solid, with long
excursions before a velocity reversing scattering from one of the
oscillators in the system. At low temperatures transport occurs via hopping
of the carrier between different unit cells, with the particle spending a
long time in the neighborhood of a given oscillator before making a
transition to a neighboring one. In this hopping limit the diffusion
constant for untrapped particles varies as $T^{3/4},$ ultimately vanishing
at zero temperature. An ensemble of particles in thermal equilibrium with
the chain contains both itnerant and self-trapped populations of particles,
and exhibits a diffusion constant $D\sim T^{1/2}e^{-T_{0}/T}$ that is
exponentially activated at low temperatures, with a functional form similar
to that derived for polaron hopping in quantum mechanical treatments of the
problem.

In the current classical model, we take the particle-oscillator interaction
to be linear in the oscillator coordinate when the particle finds itself
within a distance $\sigma <a/2$ of the oscillator. Thus, the line on which
the particle moves naturally decomposes into unit cells of width $a$, each
centered on an interacting subcell of width $2\sigma .$ This interacting
subcell is surrounded on each side by noninteracting regions of width $%
a-2\sigma =L$. (See Fig.1.) The total Hamiltonian describing this system can
be written\thinspace \cite{footnote1} 
\begin{equation}
H=\frac{1}{2}p^{2}+\sum_{m}\frac{1}{2}\left( \pi _{m}^{2}+\omega ^{2}\phi
_{m}^{2}\right) +\alpha \sum_{m}\phi _{m}n_{m}\left( q\right)  \label{Ham}
\end{equation}%
where $p$ and $q$ are the momentum and position of the particle, and $\pi
_{m}$ and $\phi _{m}$ the momentum and displacement of the $m^{th}$
oscillator, which is located on the chain at $q=q_{m}=ma.$ In the
Hamiltonian (\ref{Ham}), the function $n_{m}\left( q\right) =\theta \left(
\sigma -\left| q-q_{m}\right| \right) $ governs the range of interaction
between the particle and the $m$th oscillator: it vanishes outside the
interaction region associated with that oscillator and is equal to unity
inside it. We denote by $\theta \left( x\right) $ the unit increasing step
function. The parameter $\alpha $ governs the strength of the interaction.

Although not necessary, it is convenient to think of the oscillator motion
in this model as occurring in a direction perpendicular to that of the
particle, as depicted in Fig.\thinspace \ref{FIG1}. At any given instant the
interaction term defines a disordered step potential $U(q) = \alpha
\sum_{m}\phi _{m}n_{m}(q)$ seen by the particle as it moves along the chain.
Under the classical evolution of the model, the vibrational modes in
subcells in which the particle is absent (i.e., modes for which $n_{m}\left(
q\right) =0$) oscillate independently around their non-interacting
equilibrium position $\phi_{m} =0.$ During times that the particle is in one
of the interacting regions, however, the associated vibrational mode in that
region oscillates about a displaced equilibrium position $\phi _{\ast
}=-\alpha /\omega ^{2}$. This local displacement of the equilibrium position
of the oscillator in the region around the particle can be viewed as a
vibrational distortion of the medium that ``accompanies'' the particle as it
moves, forming part of this classical polaron. Indeed, the ground state of
this model corresponds to a polaronic ``self-trapped'' state in which the
particle is at rest in one of the interacting regions, with all oscillators
at rest in their normal or displaced equilibrium positions. For this state,
the total energy of the system%
\begin{equation}
E_{g}=-\frac{1}{2}\omega ^{2}\phi _{\ast }^{2}=-\frac{\alpha ^{2}}{2\omega
^{2}}=-E_{B}  \label{Ebind}
\end{equation}%
is negative and has a magnitude $E_{B}=\alpha ^{2}/2\omega ^{2},$ analogous
to what is referred to in the polaron literature as the ``polaron binding
energy''. Thus, the binding energy $E_{B}$ provides a natural measure of the
coupling between the particle and the local oscillator modes.

In our classical model, except for instances when the particle is at the
boundary between an interacting and non-interacting region, the particle and
oscillator motion are decoupled and evolve independently, with the particle
moving at constant speed. At ``impacts'', i.e., at moments when the particle
arrives at the boundary between a non-interacting region and an interacting
region, the particle encounters a discontinuity $\Delta U\equiv \Delta =\pm
\alpha \phi _{m}$ in the potential energy, and thus an impulsive force, that
is determined by the oscillator displacement in the interacting region at
the time of impact. When the particle energy $\varepsilon $ at the moment of
impact is less than $\Delta ,$ the particle is reflected back into the
subcell in which it was initially moving; when $\varepsilon >\Delta $ the
particle moves into the next subcell, with a new speed and energy $%
\varepsilon _{f}=\varepsilon -\Delta $ governed by energy conservation.
Thus, the general motion of the particle along the oscillator chain involves
sojourns within each interacting or noninteracting subcell it visits, during
which it can reflect multiple times, interspersed with impact events in
which the particle's energy and the oscillator displacements allow it to
pass into a neighboring cell.

This uncoupled evolution of the two subsystems between impacts makes it easy
to perform a numerical ``event driven'' evolution of the system by focusing
on the changes that occur during impacts. We have performed a large number
of such numerical evolutions of this model for a set of initial conditions
in which the particle is injected into the chain at its midpoint, with each
oscillator initialized to a state drawn independently and randomly from a
thermal distribution at temperature $T$. Our numerical studies, which are
supported by theoretical analysis given below, suggest that for such initial
states there are three distinct types of dynamical behavior that can occur.
One of these is essentially trivial, and leads to localized self-trapped
particle motion only. The other two types both give rise to diffusive motion
of the particle over large time and length scales. In all cases the
oscillator distribution as a whole remains in thermal equilibrium.

The self-trapped states of the model correspond to configurations in which a
particle is located initially inside an interaction region (at $q_{m}$,
say), with the energy 
\begin{equation}
\varepsilon_{int}=\frac{1}{2}p^{2}+\frac{1}{2}\left( \pi_{m}^{2}+\omega
^{2}\phi_{m}^{2}\right) +\alpha\phi_{m}
\end{equation}
of the particle and the oscillator in that region partitioned in such a way
that the particle is unable to escape. Such states include all those bound
states in which $\varepsilon_{int}$ is negative. As shown in Ref. \cite{DPS}%
, however, it also includes a set of positive energy self-trapped states,
i.e., all dynamical states for which $E_{B}>\varepsilon_{int}>0$, and for
which the associated oscillator energy 
\begin{equation}
\varepsilon_{\text{osc}}=\frac{1}{2}\left( \pi_{m}^{2}+\omega^{2}\left(
\phi_{m}-\phi_{\ast}\right) ^{2}\right) <E_{c}=E_{B}-\varepsilon\left( 1-%
\frac{\varepsilon}{4E_{B}}\right)  \label{SelfTrapped}
\end{equation}
is sufficiently low that the potential barrier seen by the particle at each
impact is always greater than its kinetic energy. Clearly for all such
initial conditions the mean square displacement $\langle q^{2}\left(
t\right) \rangle=\langle m^{2}\left( t\right) \rangle a^{2}$ of the
particle, measured (or course-grained) in terms of the mean square number $%
\langle m^{2}\left( t\right) \rangle$ of unit cells visited, remains equal
to zero.

\begin{figure}[t]
\center \includegraphics[height=2.5in]{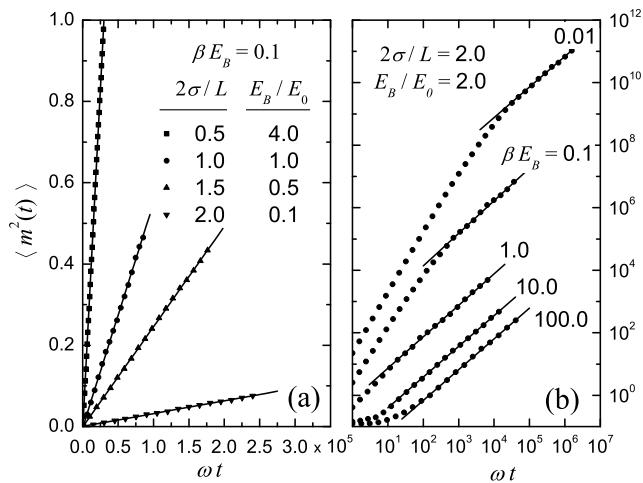}
\caption{Mean square displacement (measured in units of the squared cell
spacing $a^{2}$) for an ensemble of particles injected with positive energy
into an oscillator chain in thermal equilibrium at temperature $T=1/k\protect%
\beta$, with temperatures, coupling strengths, and mechanical parameters as
indicated. For the parameters in the linear plot on the left $\langle
m^{2}\left( t\right) \rangle$ appears linear over the entire range of times.
On the log-log plot on the right one can see that linear growth of $\langle
m^{2}\left( t\right) \rangle$ begins after an initial period of super- or
sub- diffusive behavior. The temperature range in the latter figure covers
both the high temperature and low temperature regimes analyzed in the text.}
\label{FIG2}
\end{figure}

Our numerical studies indicate that for initial states in which the particle
is \emph{not} in a self-trapped state, the particle at sufficiently long
times undergoes diffusive motion, with a mean square displacement $\langle
q^{2}\left( t\right) \rangle \sim Dt$ that grows linearly in time. Figure %
\ref{FIG2} shows a representative set of numerical data for the
(dimensionless) mean square displacement $\langle m^{2}\left( t\right)
\rangle =\langle q^{2}\left( t\right) \rangle /a^{2}$ for several different
parameters and temperatures $T$, the latter expressed in terms of the
reciprocal temperature $\beta =1/kT,$ where $k$ is Boltzmann's constant, and
with each curve corresponding to particle evolutions (or trajectories)
averaged over $N_{t}=10^{5}$ independent initial conditions \cite{footnote2}%
. In the numerical calculations presented here, time is expressed in terms
of the oscillator frequency through the dimensionless combination $\tau
=\omega t,$ lengths are measured in terms of the width $2\sigma $ of the
interaction region, and energies are measured in terms of the quantity 
\begin{equation}
E_{0}=\sigma ^{2}\omega ^{2},
\end{equation}%
which is one-half the kinetic energy of a particle that traverses an
interaction region in time $t=\omega ^{-1}.$ Thus, aside from the
temperature, there are only two independent mechanical parameters in the
problem, which we can take to be the ratio $\eta =2\sigma /L$ of the width
of the interacting and noninteracting regions, and the ratio $\varepsilon
=E_{B}/E_{0}=\alpha ^{2}/2\sigma ^{2}\omega ^{4}$ of the polaron binding
energy to $E_{0}$.

\begin{figure}[t]
\center \includegraphics[height=2.5in]{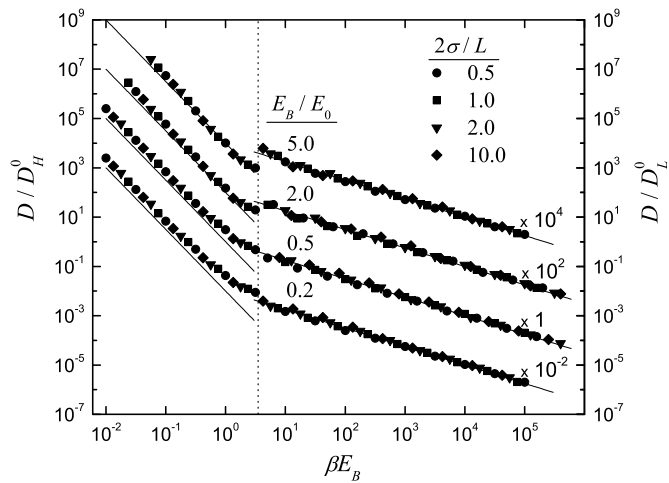}
\caption{Diffusion constant obtained from linear fits to long-time mean
square displacement, for a wide range of system parameters and coupling
strengths as indicated, as a function of the inverse temperature $\protect%
\beta E_{B}.$ Curves corresponding to different coupling strengths are
multiplied by different powers of 10 as indicated for clarity. Data on the
left (right) hand side has been scaled by the high (low) temperature value $%
D_{H}^{0}$ ($D_{L}^{0}$) derived in the text. Straight lines are the power
laws $(\protect\beta E_{B})^{5/2}$ for the high temperature data and $(%
\protect\beta E_{B})^{3/4}$ for the low temperature data.}
\label{FIG3}
\end{figure}

The diffusion constant $D$ describing the linear increase in the mean square
displacement of initially untrapped particles, obtained from a fit to the
numerical data for $\langle q^{2}\left( t\right) \rangle $ in the region
where it becomes linear, is found to be a strong function of the mechanical
parameters and of the temperature. Results for a range of parameters and a
wide range of temperature are presented in Fig. \ref{FIG3}. Note that data
for different values of the coupling strength parameter $E_{B}/E_{0}$ are
shifted by powers of ten to separate the data. If this scaling were not done
the data would all lie on a single pair of universal curves. As the figure
shows, for any given fixed set of parameters, the temperature dependence of
the diffusion constant takes two simple limiting forms. At high
temperatures, the dependence of the diffusion constant on temperature
follows a power law%
\begin{equation}
D/D_{H}^{0}\sim \left( \beta E_{B}\right) ^{-5/2}  \label{HT}
\end{equation}%
where $D_{H}^{0}$ is a constant that depends on the mechanical parameters.
In Sec. \ref{highT} we derive an approximate expression for $D_{H}^{0}$ [See
Eq. (\ref{DH0})] which we have used to scale the data on the left hand side
of the figure. The straight lines in the left hand side of the figure are
the function $\left( \beta E_{B}\right) ^{-5/2}.$

Also as seen in Fig. \ref{FIG3}, as the temperature is lowered the diffusion
constant crosses over to a weaker power law of the form%
\begin{equation}
D/D_{L}^{0}\sim \left( \beta E_{B}\right) ^{-3/4}  \label{LT}
\end{equation}%
where $D_{L}^{0}$ is a different constant, which we derive in Sec. \ref{lowT}%
, below and which we have used to scale the data on the right hand side of
that figure. The straight lines in the right hand side of that figure are
the function $\left( \beta E_{B}\right) ^{-3/4}$.

These two different power laws result from two quite different forms of
microscopic dynamics that, in a sense, correspond to the adiabatic and
nonadiabatic polaron dynamics studied in quantum mechanical treatments of
the problem. Indeed, at sufficiently high temperatures the characteristic
particle speed $p\sim\sqrt{kT}$ becomes sufficiently large that the time for
a particle to traverse the interaction region is short compared to the
oscillator period. In this limit, also, the typical potential energy barrier 
$\Delta\sim \sqrt{2E_{B}kT}$ encountered by the particle at impacts becomes
small compared to the particle energy. The picture that emerges is that, at
high temperatures, a particle with a (high) initial speed, close to the
thermal average, will pass through many interaction regions in succession,
typically losing a small amount of energy to each oscillator as it does so.
Eventually, this continual loss of energy, which acts like a source of
friction, slows the particle down to a speed where it is more likely to
encounter a barrier larger than its energy, at which point it undergoes a
velocity reversing (or randomizing) kick back up to thermal velocities. By
heuristically taking into account a typical excursion $\ell\left( p\right) $
of a particle with initial momentum $p$ we derive in Sec. \ref{highT} the
result (\ref{HT}) seen in our numerical data, and in the process derive an
approximate expression for the scaling prefactor $D_{H}^{0}$. This regime is
similar to the so-called adiabatic limit discussed in the small polaron
literature, in which the time for a bare particle to move from one cell to
the next is short compared to the evolution time of the oscillator. Indeed,
in this limit the random potential seen by the particle typically changes
adiabatically with respect to the particle's net motion.

As the temperature is lowered, however, the mean particle speed decreases.
At low enough temperatures a typical particle traverses the interaction
region in a time that is large compared to the oscillator period, so that at
each impact the oscillator phase is randomized. In this limit, also, the
typical barrier energy $\sqrt{2E_{B}kT}$ becomes large compared to the
particle energy. As a result, a particle will typically undergo many
reflections inside the interaction region before escaping into one of the
adjacent regions. The resulting motion can be described as a random walk
between neighboring unit cells, with the diffusion constant depending on the
escape rate from a given cell. In this low-temperature hopping limit, the
motion is similar to the nonadiabatic limit of small polaron motion, in
which the time to move freely from one cell to another is very long compared
to the typical oscillator period. In Sec. \ref{lowT} we calculate this
escape rate in order to produce a theoretical estimate of the diffusion
constant in the low temperature nonadiabatic limit, represented by the solid
lines on the right hand side of Fig. \ref{FIG3}.

In Sec. \ref{thermal} we combine the results of the two analyses in the
preceding sections to show that the behavior of an ensemble of particles in
thermodynamic equilibrium on such a chain would exhibit diffusive behavior
that retains all essential features of the high temperature limit, but
becomes exponentially suppressed in the low temperature hopping limit as a
result of the increasing fraction of self-trapped particles at low
temperatures. This results in a thermally activated diffusion constant very
similar to that derived in the nonadiabatic limit using quantum mechanical
arguments. The last section contains a summary and discussion.

\section{Diffusion of Injected Carriers at High Temperatures}

\label{highT}

In this section we consider the dynamical behavior of this model at high
temperatures, i.e., the left hand side of Fig. (\ref{FIG3}). We note first
that when a particle which is initially in one of the non-interacting
regions attempts to enter an interacting region, the distribution of
potential energy steps $\Delta =\alpha \phi _{m}$ it encounters is given by
the relation%
\begin{equation}
\rho \left( \Delta \right) =\sqrt{\frac{\beta }{4\pi E_{B}}}e^{-\beta \Delta
^{2}/4E_{B}},
\end{equation}%
which follows from the assumed thermal distribution $\rho \left( \phi
_{m}\right) =\sqrt{\beta \omega ^{2}/2\pi }e^{-\beta \omega ^{2}\phi
_{m}^{2}/2}$ for the displacement $\phi _{m}$ of the oscillator at the
moment of impact and the definition (\ref{Ebind}). Consequently, the
magnitude of the typical (RMS) potential step 
\begin{equation}
\Delta =\sqrt{2E_{B}kT}
\end{equation}%
seen by the particle, independent of its energy, increases only as the
square root of the temperature. At asymptotically high temperatures the
typical step is, therefore, relatively small compared to the energy $%
\varepsilon \sim kT/2$ of the particle. Thermalized particles will therefore
easily pass through many cells in the same direction with high probability.
In this limit the traversal time becomes short compared to the oscillator
period, and the particle will tend to lose energy to the oscillator bath at
a well defined average rate. To see this, we consider a particle with an
initially high initial momentum $p$ which has just entered the interaction
region associated with one of the oscillators in the chain. Before impact,
the oscillator in that region is (by assumption) in thermal equilibrium
about the undisplaced equilibrium position $\left( \phi =0\right) $, and so
the average energy step $\langle \Delta _{0}\rangle =\alpha \langle \phi
_{0}\rangle $ encountered by the particle at the interface vanishes. After
entry, the state of the oscillator in the interaction region is
characterized by the distribution 
\begin{equation}
\rho \left( \phi _{0},\pi _{0}\right) =\sqrt{\frac{\beta \omega ^{2}}{2\pi }}%
e^{-\beta \omega ^{2}\phi _{0}^{2}/2}\sqrt{\frac{\beta }{2\pi }}e^{-\beta
\pi _{0}^{2}/2}\theta \left( p^{2}-2\alpha \phi _{0}\right) ,
\label{rhoinitial}
\end{equation}%
where the upper cutoff on $\phi $ associated with the step function arises
from the requirement that the initial energy of the particle and the
oscillator state be such as to have allowed the particle to enter the
interaction region in the first place. For high temperatures (i.e., $\beta
E_{B}\ll 1$) and particles with thermal energies $p^{2}\sim kT,$ this cutoff 
$\phi _{c}\sim kT/2\alpha $ is well outside the range $\phi _{th}\sim \sqrt{%
kT}/\omega $ of the thermal distribution for the oscillator displacement and
can be neglected, i.e., $\phi _{c}/\phi _{th}\sim \left( \beta E_{B}\right)
^{-1/2}\gg 1$. Once the particle is in the interaction region, the
oscillator state, and hence the distribution $\rho \left( \phi ,\pi
,t\right) $ will evolve in time as the oscillator starts to oscillate about
its \emph{displaced} equilibrium position $\phi _{\ast }.$ After a time $%
t=2\sigma /p,$ the particle will have traversed the interaction region, and
the oscillator coordinate $\phi $ will have changed by an amount 
\begin{equation}
\Delta \phi =\left( \phi _{0}-\phi _{\ast }\right) \left( \cos \left(
2\omega \sigma /p\right) -1\right) +\frac{\pi _{0}}{\omega }\sin \left(
2\omega \sigma /p\right) .
\end{equation}%
Averaging this over the initial distribution (\ref{rhoinitial}) gives the
average change in the oscillator displacement 
\begin{equation}
\langle \Delta \phi \rangle =\phi _{\ast }\left( 1-\cos \left( 2\omega
\sigma /p\right) \right)
\end{equation}%
at the moment the particle attempts to leave the interaction region after
one pass. When it does so, the average difference between the energy steps
that the particle crosses between entering and leaving the interaction
region must equal the average change $\langle \Delta \varepsilon \rangle $
in the particle's energy as it passes through that region, i.e., 
\begin{equation}
\langle \Delta \varepsilon \rangle =-\alpha \langle \Delta \phi \rangle
=-2E_{B}\left[ 1-\cos \left( 2\omega \sigma /p\right) \right] .
\label{deltaE}
\end{equation}%
Of course, at temperatures high enough that $\sqrt{kT}\gg 2\sigma \omega $,
the oscillator period will be large compared to the characteristic time for
particles of average thermal energy to traverse the interaction region. For
such particles, we can expand the cosine in (\ref{deltaE}) to find that the
average energy change 
\begin{equation}
\langle \Delta \varepsilon \rangle =-4E_{B}E_{0}/p^{2}
\end{equation}%
is negative, simply because the particle exerts a constant force on the
oscillator in the interaction region through which it is passes that tends
to decrease the potential energy in that region while it does so, and as a
result makes it more likely for the particle to lose net kinetic energy when
it passes back out into the noninteracting region. Thus the particle tends
to slow down in time as it passes through many such cells in succession. The
average momentum loss 
\begin{equation}
\frac{dp}{dn}=\frac{dp}{d\varepsilon }\frac{\langle \Delta \varepsilon
\rangle }{dn}=-\frac{4E_{B}E_{0}}{p^{3}},  \label{dpdn}
\end{equation}%
per cell traversed can then be integrated to obtain the typical distance%
\begin{equation}
\ell \left( p\right) =an\left( p\right) =\frac{p^{4}a}{16E_{B}E_{0}}
\label{l(p)}
\end{equation}%
the particle must travel, losing energy at this rate, to dissipate most of
its initial momentum $p$. This is, in fact, an overestimate since the
particle need only dissipate as much momentum is required for it to
encounter, with significant probability, a barrier larger than its (reduced)
energy. Of course, in the high temperature limit, when barriers of
sufficient energy are few, the particle will need to dissipate a significant
fraction of its initial energy. We also note from (\ref{dpdn}) that for fast
enough particles the fractional change in momentum per traversal is small,
so the average time to traverse each complete cell can be written $%
dt/dn=a/p. $ This allows us to integrate the average ``frictional force''%
\begin{equation}
\frac{dp}{dt}=\frac{dp}{dn}\frac{dn}{dt}=-\frac{4E_{B}E_{0}}{p^{2}a}
\end{equation}%
on particles with large initial momentum to obtain the characteristic time%
\begin{equation}
\tau \left( p\right) =\frac{p^{3}a}{12E_{B}E_{0}}  \label{tau(p)}
\end{equation}%
for this slowing down to occur.

The picture that develops from this analysis, therefore, is that,
fluctuations aside, at high temperature a thermal distribution of particles
will contain many particles of high energy that are slowing down on average
in response to a frictional force arising from their interaction with the
oscillators through which they pass. This slowing down can not continue,
however, or else the distribution would not maintain itself in thermal
equilibrium. Indeed, once the particle speed is slow enough that it becomes
likely to encounter a barrier larger than its kinetic energy it receives a
velocity reversing impulse. After such an event, if the time for the
particle to traverse the interaction region is still small compared to the
oscillator period the particle will continue to lose energy, on average,
until its traversal time becomes comparable to the oscillator period, in
which limit it can gain as well as lose energy due to the random phases of
the oscillator at the moment it attempts to leave the interaction region. In
the overall process a particle with a given initial momentum $p$ will travel
a distance $\ell \left( p\right) $ in a time $\tau \left( p\right) $ before
ultimately being kicked back up to higher energy. The associated diffusion
constant can, we argue, be calculated as in a random walk with a
distribution of hopping times $\tau $ and step lengths $\ell $, i.e., 
\begin{equation}
D=2\int_{0}^{\infty }dp\;\rho (p)\;\frac{\ell ^{2}(p)}{\tau \left( p\right) }
\label{Dflight}
\end{equation}%
where $\rho \left( p\right) =\sqrt{\beta /2\pi }e^{-\beta p^{2}/2}$ is the
thermal distribution of particle speeds. Assuming that the main contribution
to the growth of the mean square displacement comes from the fast particles
with long excursion lengths, as described above, we combine Eqs. (\ref{l(p)}%
), (\ref{tau(p)}), and (\ref{Dflight}) to write the diffusion constant in
the high temperature limit as 
\begin{equation}
D=\frac{3a}{32E_{B}E_{0}}\sqrt{\frac{\beta }{2\pi }}\int_{0}^{\infty
}p^{5}e^{-\beta p^{2}/2}dp=D_{H}^{0}\cdot \left( \beta E_{B}\right) ^{-5/2}
\end{equation}%
where%
\begin{equation}
D_{H}^{0}=\sqrt{\frac{9E_{B}a^{2}}{32\pi }}\frac{E_{B}}{E_{0}}.  \label{DH0}
\end{equation}%
\qquad This expression for the diffusion constant is represented as the
solid lines in the left hand side of Fig. \ref{FIG3}. Agreement of the
numerical data with the scaling behavior predicted above is excellent. This
collapse of the data for a wide range of values of $E_{B}/E_{0}$ and $%
2\sigma /L$ suggests that the heuristic arguments given above capture the
essential physics taking place in the high temperature limit, even though
they lead to an estimate of $D_{H}^{0}$ that is too low by a numerical
factor lying between 1 and 2.5 over the range of parameters studied.

\section{Diffusion of Injected Carriers at Low Temperatures}

\label{lowT}

In this section we provide an analysis aimed at understanding the diffusion
constant in the system at very low temperatures, i.e., the right hand side
of Fig. (\ref{FIG3}). As we have already argued, in this limit the barrier
seen by a particle in a given subcell can be large compared to typical
particle energies. Consequently, an injected particle that makes its way
into an interaction region can rattle around in that region, undergoing many
reflections before encountering a barrier sufficiently low that it is able
to pass back out. Also, as the temperature is lowered, the typical time for
a thermal particle to traverse the interaction region becomes very large
compared to the oscillator period. Hence, after each traversal, the phase of
the oscillator is randomized, except for a set of velocities of zero measure
for which the traversal time is commensurate with the oscillator period.
Thus, the motion of the particle at very low temperatures becomes a random
walk between neighboring cells, and calculating the diffusion constant 
\begin{equation}
D\sim\Gamma a^{2}  \label{Dgamma}
\end{equation}
reduces to a calculation of the average total escape rate $\Gamma$ for
leaving an individual unit cell.

It is relatively easy to see, also, that at low enough temperatures the time
for the particle to leave a given unit cell is dominated by the time for
leaving the interaction region at its center. Intuitively this is clear,
since as a thermal particle approaches an interface from the non-interacting
side, the oscillator displacement in the interaction region is equally
likely to be positive or negative, but with a magnitude $\Delta \sim \sqrt{%
E_{B}kT}$ that is large relative to the particle energy $\varepsilon \sim
kT/2$. Hence, at low temperatures the probability for the particle to pass 
\emph{into} the interaction region at any given impact approaches one-half.
Once inside the interaction region, however, the oscillator therein starts
oscillating about its displaced equilibrium position. The initial amplitude
of the oscillator at the moment the particle enters will typically be
negative (so the particle could get in) and with a magnitude $\left| \phi
_{0}\right| \sim \sqrt{kT}/\omega $ that is small compared to the magnitude
of the \emph{maximum} displacement $\phi _{\text{max}}\sim -2\phi _{\ast
}+\left| \phi _{0}\right| \sim -2\phi _{\ast }$ of the oscillations that it
will now perform about its suddenly shifted equilibrium position. If, after
the particle traverses the interaction region, the phase of the oscillator
has become randomized, it is then very \emph{unlikely} that the amplitude
will allow the particle to exit the interaction region, since for most of
the oscillation the amplitude is a significant fraction of $\phi _{\text{max}%
}$, and thus will present a barrier $\Delta $ which is a significant
fraction of the maximum barrier $\Delta _{\text{max}}\sim 2\alpha \phi
_{\ast }\sim 4E_{B},$ which in the low temperature limit $\beta E_{B}\gg 1$
is large relative to the particle energy. Consequently, in this limit the
particle will typically reflect many times inside the interaction region
until it arrives at the interface at a moment when the oscillator phase is
very close to its value at the moment it entered.

Calculation of the diffusion constant in the low temperature limit,
therefore, reduces to a calculation of the average escape rate for a
particle which has made its way into the interaction region from the
outside. To this end, we focus on that subensemble of particles which have
just entered an interaction region, and which end up with energy $%
\varepsilon =p^{2}/2$ once inside the region. As we will see, the
conditional distribution $P\left( \phi ,\pi |\;\varepsilon \right) $ of
oscillator states at the moment a particle enters the interaction region
with this energy is straightforward to calculate, and allows us to obtain
the corresponding conditional distribution $P\left( \varepsilon _{\text{osc}%
}|\,\varepsilon \right) $ of oscillator energies. Assuming that the phase of
the oscillator is completely randomized as the particle traverses the
interaction region, we can then use this to compute the distribution $\rho
\left( \Delta |\;\varepsilon \right) $ of potential energy barriers $\Delta
=\alpha \phi $ that a particle with this energy will see as it attempts to
leave the region. It will do so if the barrier $\Delta $ is less than the
energy $\varepsilon $ of the particle, so the a priori probability $Q\left(
\varepsilon \right) $ that a particle which has entered the interaction
region with energy $\varepsilon $ will exit the region after any given
impact is 
\begin{equation}
Q\left( \varepsilon \right) =\int_{-\infty }^{\varepsilon }\rho _{\text{osc}%
}\left( \Delta \;|\,\varepsilon \right) \;d\Delta .
\end{equation}%
The average rate at which a particle of this energy leaves the interaction
region can then be written as the product 
\begin{equation}
\gamma \left( \varepsilon \right) =Q\left( \varepsilon \right) \frac{p}{%
2\sigma }  \label{gamma}
\end{equation}%
of the attempt frequency $\nu =p\left( \varepsilon \right) /2\sigma $ and
the success probability $Q\left( \varepsilon \right) $. The total average
rate $\Gamma $ at which particles leave the interaction region can finally
be evaluated%
\begin{equation}
\Gamma =\int_{0}^{\infty }\;\rho _{u}^{\text{int}}\left( \varepsilon \right)
\gamma \left( \varepsilon \right) d\varepsilon  \label{GAMMA}
\end{equation}%
using the the \emph{steady state} distribution $\rho _{u}^{\text{int}}\left(
\varepsilon \right) $ of untrapped particles in the interaction region.

We focus initially, therefore, on particles just about to enter the
interaction region, with external initial energy $\varepsilon _{\text{out}},$
which is assumed to be thermally distributed, $\rho \left( \varepsilon _{%
\text{out}}\right) =\sqrt{\beta /\pi \varepsilon _{\text{out}}}\exp \left(
-\beta \varepsilon _{\text{out}}\right) $, as is the state of the oscillator
in the interaction region it is attempting to enter. Under these
circumstances, the particle can only enter the interaction region if $%
\varepsilon _{\text{out}}$ is larger than the potential step it encounters,
i.e., if $\varepsilon _{\text{out}}-\alpha \phi =\varepsilon $ is positive,
where $\varepsilon $ is the energy of the particle once it has entered the
interaction region. Up to a normalization constant, therefore, the joint
distribution of the oscillator state and the particle energy at the moment
it enters the interaction region is 
\begin{equation}
f\left( \pi ,\phi ,\varepsilon \right) =\sqrt{\frac{\beta }{\pi \left(
\varepsilon +\alpha \phi \right) }}\frac{\beta \omega }{2\pi }e^{-\beta
\left( \omega ^{2}\phi ^{2}+\pi ^{2}\right) /2}e^{-\beta \left( \varepsilon
+\alpha \phi \right) }\theta \left( \varepsilon +\alpha \phi \right)
\end{equation}%
where the step function ensures that the original particle energy and
oscillator configuration were such as to allow it to enter in the first
place. The corresponding conditional distribution of oscillator states in an
interaction region into which a particle of energy $\varepsilon $ has just
passed is then%
\begin{equation}
P\left( \pi ,\phi |\;\varepsilon \right) =\frac{1}{Z\left( \varepsilon
\right) }\sqrt{\frac{\beta }{\pi \left( \varepsilon +\alpha \phi \right) }}%
e^{-\beta \left( \omega ^{2}\phi ^{2}+\pi ^{2}\right) /2}e^{-\beta \left(
\varepsilon +\alpha \phi \right) }\theta \left( \varepsilon +\alpha \phi
\right)  \label{P5}
\end{equation}%
where%
\begin{equation}
Z\left( \varepsilon \right) =\int_{-\infty }^{\infty }d\phi \int_{-\infty
}^{\infty }d\pi \;f\left( \pi ,\phi ,\varepsilon \right) =\frac{1}{\omega }%
\sqrt{1-\frac{\varepsilon }{2E_{B}}}e^{-\beta \left( \varepsilon
-E_{B}\right) }e^{-\kappa ^{2}/2}K_{1/4}\left( \kappa ^{2}/2\right) .
\end{equation}%
In this last expression, $K_{\nu }\left( z\right) $ is the modified Bessel
function, and we have introduced the quantity 
\begin{equation}
\kappa \left( \varepsilon \right) =\sqrt{\beta E_{B}}\left( 1-\frac{%
\varepsilon }{2E_{B}}\right) .
\end{equation}%
With the particle in the interaction region, the total mechanical energy of
the oscillator as it oscillates about its displaced equilibrium position can
be written 
\begin{equation}
\varepsilon _{\text{osc}}=\frac{1}{2}\left[ \pi ^{2}+\omega ^{2}\left( \phi
-\phi _{\ast }\right) ^{2}\right] \equiv \frac{1}{2}\left[ \pi ^{2}+\omega
^{2}\tilde{\phi}^{2}\right] .
\end{equation}%
The conditional distribution for the oscillator energy can be computed from (%
\ref{P5}) using the relation 
\begin{equation}
P\left( \varepsilon _{\text{osc}}|\;\varepsilon \right) =\int_{-\infty
}^{\infty }d\phi \int_{-\infty }^{\infty }d\pi \;P\left( \pi ,\phi
|\;\varepsilon \right) \delta \left( \varepsilon _{\text{osc}}-\frac{1}{2}%
\left[ \pi ^{2}+\omega ^{2}\tilde{\phi}^{2}\right] \right) .  \label{P6}
\end{equation}%
We find, after some work, that%
\begin{equation}
P\left( \varepsilon _{\text{osc}}|\;\varepsilon \right) =\left\{ 
\begin{array}{ccc}
0 &  & \varepsilon <\varepsilon _{-} \\ 
A\left( \varepsilon _{\text{osc}},\varepsilon \right) K\left[ \sqrt{\mu
\left( \varepsilon _{\text{osc}},\varepsilon \right) }\right] &  & 
\varepsilon _{-}<\varepsilon <\varepsilon _{+} \\ 
\sqrt{\mu \left( \varepsilon _{\text{osc}},\varepsilon \right) }A\left(
\varepsilon _{\text{osc}},\varepsilon \right) K\left[ \sqrt{\mu ^{-1}\left(
\varepsilon _{\text{osc}},\varepsilon \right) }\right] &  & \varepsilon
_{+}<\varepsilon%
\end{array}%
\right.  \label{P7}
\end{equation}%
where $K\left( m\right) $ is the complete elliptic integral of the first
kind, $\varepsilon _{\pm }=2\left( E_{B}\pm \sqrt{\varepsilon _{\text{osc}%
}E_{B}}\right) $, and we have introduced the functions%
\begin{equation}
A\left( \varepsilon _{\text{osc}},\varepsilon \right) \equiv \frac{\beta
^{3/2}}{\pi ^{3/2}\left( \varepsilon _{\text{osc}}E_{B}\right) ^{1/4}Z\left(
\varepsilon \right) }e^{-\beta \left( \varepsilon +\varepsilon _{\text{osc}%
}-E_{B}\right) }
\end{equation}%
and%
\begin{equation}
\mu \left( \varepsilon _{\text{osc}},\varepsilon \right) =\frac{\sqrt{%
\varepsilon _{\text{osc}}/E_{B}}+\varepsilon /2E_{B}-1}{2\sqrt{\varepsilon _{%
\text{osc}}/E_{B}}}.
\end{equation}%
As we will see below, at very low temperatures the distribution $\rho _{u}^{%
\text{int}}\left( \varepsilon \right) $ of untrapped particles in Eq. (\ref%
{GAMMA}), becomes exponentially suppressed at higher energies, and it
suffices to consider the form of $P\left( \varepsilon _{\text{osc}%
}|\varepsilon \right) $ for values of $\varepsilon $ less than or of the
order of a few $kT\ll E_{B}$. For such values of $\varepsilon ,$ the
function $P\left( \varepsilon _{\text{osc}}|\;\varepsilon \right) ,$ as a
function of $\varepsilon _{\text{osc}}$, becomes strongly peaked in the
neighborhood of $\varepsilon _{\text{osc}}\sim E_{B}$. This can be seen from
Eq.(\ref{P7}), where it is clear that for values of $\varepsilon $ in this
range the condition for validity of the last line is never satisfied. The
first line then shows that $P\left( \varepsilon _{\text{osc}}|\;\varepsilon
\right) $ vanishes except for oscillator energies $\varepsilon _{\text{osc}}$
satisfying%
\begin{equation}
\varepsilon _{\text{osc}}\geq E_{B}-\varepsilon \left( 1-\frac{\varepsilon }{%
4E_{B}}\right) \sim E_{B}\left( 1-\frac{1}{\beta E_{B}}\right)
\end{equation}%
which is very close to, and just below, $E_{B}$ at low temperatures.
Moreover the function $A\left( \varepsilon _{\text{osc}},\varepsilon \right) 
$ drives the distribution $P\left( \varepsilon _{\text{osc}}|\;\varepsilon
\right) $ exponentially towards zero for increasing oscillator energies
greater than a few $kT$ of where it first becomes non-zero. Thus, at low
temperatures, for values of $\varepsilon $ of interest, the distribution $%
P\left( \varepsilon _{\text{osc}}|\;\varepsilon \right) $ becomes
increasingly peaked in the variable $\varepsilon _{\text{osc}}$ in the
neighborhood of $E_{B}.$ Since the function $\mu \left( \varepsilon _{\text{%
osc}},\varepsilon \right) \sim \varepsilon /4E_{B}$ is small in this region,
we can approximate the elliptic integral in (\ref{P7}) by its limiting value 
$K\left( 0\right) =\pi /2$. After some analysis, we find that at low
temperatures the distribution $P\left( \varepsilon _{\text{osc}%
}|\;\varepsilon \right) $ asymptotically approaches the limiting distribution%
\begin{equation}
P\left( \varepsilon _{\text{osc}}|\;\varepsilon \right) \sim \frac{e^{-\beta
\varepsilon _{\text{osc}}}\ }{\left( \varepsilon _{\text{osc}}\right)
^{1/4}Z^{\prime }}\theta \left[ \varepsilon -2\left( E_{B}-\sqrt{\varepsilon
_{\text{osc}}E_{B}}\right) \right]  \label{PEoscEin}
\end{equation}%
where%
\begin{equation}
Z^{\prime }=\frac{\Gamma \left( 3/4,\kappa ^{2}\right) }{\beta ^{3/4}}\sim 
\frac{\text{exp}\left( -\beta E_{B}\left( 1-\frac{\varepsilon }{2E_{B}}%
\right) ^{2}\right) }{\beta ^{3/4}\left( \beta E_{B}\left( 1-\frac{%
\varepsilon }{2E_{B}}\right) ^{2}\right) ^{1/4}},\qquad \beta E_{B}\gg 1,
\label{Z}
\end{equation}%
in which $\Gamma \left( a,z\right) $ is the incomplete gamma function, the
asymptotic form of which we have used in the second expression.

Now for a given oscillator energy $\varepsilon _{\text{osc}}$, and assuming
uniformly distributed phases, the conditional distribution for the
displacement $\phi $ of the oscillator about its displaced equilibrium
position $\phi _{\ast },$ given the oscillator energy $\varepsilon _{\text{%
osc}}$ is easily computed. Using this, the distribution of barriers $\Delta
=-\alpha \phi $ encountered by a particle after traversing an interaction
region in which there is an oscillator of energy $\varepsilon _{\text{osc}}$
is readily found to be%
\begin{equation}
\rho _{B}\left( \Delta |\;\varepsilon _{\text{osc}}\right) =\frac{\theta
\left( \varepsilon _{\text{osc}}-\frac{\omega ^{2}}{2}\left( \frac{\Delta }{%
\alpha }+\phi _{\ast }\right) ^{2}\right) }{2\pi \sqrt{E_{B}\left(
\varepsilon _{\text{osc}}-\frac{\omega ^{2}}{2}\left( \frac{\Delta }{\alpha }%
+\phi _{\ast }\right) ^{2}\right) }}.  \label{rhoDeltaEosc}
\end{equation}%
Multiplying (\ref{PEoscEin}) by (\ref{rhoDeltaEosc}) and integrating over
all oscillator energies allows us to obtain the conditional barrier
distribution%
\begin{equation}
\rho \left( \Delta |\;\varepsilon \right) =\int_{\varepsilon _{1}}^{\infty
}d\varepsilon _{\text{osc}}\ P\left( \varepsilon _{\text{osc}}|\varepsilon
\right) \rho _{B}\left( \Delta |\varepsilon _{\text{osc}}\right)
\label{rhodeltaeps}
\end{equation}%
where, due to the step functions in (\ref{PEoscEin}) and (\ref{rhoDeltaEosc}%
), the lower limit of integration 
\begin{align}
\varepsilon _{1}& =\lambda E_{B} \\
\lambda & =\max \left[ \left( 1-\frac{\varepsilon }{2E_{B}}\right)
^{2},\left( \frac{\Delta }{2E_{B}}-1\right) ^{2}\right]
\end{align}%
in (\ref{rhodeltaeps}) is a function of $\varepsilon $ and $\Delta .$ As we
have already observed, at low temperature the integrand in (\ref{rhodeltaeps}%
) dies away exponentially as a function of $\varepsilon _{\text{osc}}$ from
the maximum value it takes on at the lower limit. By expanding to lowest
order the non-exponential part of the integrand about the lower limit of
integration and performing the resulting integral we find that for $\beta
E_{B}\gg 1$ and for values of $\varepsilon <2E_{B},$ the function $\rho
\left( \Delta |\;\varepsilon \right) $ is well approximated by the expression%
\begin{equation}
\rho \left( \Delta |\;\varepsilon \right) =\left\{ 
\begin{array}{ccc}
\frac{1}{2\pi E_{B}}\left[ \left( 1-\frac{\varepsilon }{2E_{B}}\right)
^{2}-\left( \frac{\Delta }{2E_{B}}-1\right) ^{2}\right] ^{-1/2} &  & \qquad 
\text{for }\left( 1-\frac{\varepsilon }{2E_{B}}\right) ^{2}>\left( \frac{%
\Delta }{2E_{B}}-1\right) ^{2} \\ 
\sqrt{\frac{\beta }{4\pi E_{B}}}\frac{\sqrt{2E_{B}-\varepsilon }}{\left|
\Delta -2E_{B}\right| ^{1/2}}e^{-\beta E_{B}\left( 1-\frac{\varepsilon }{%
2E_{B}}\right) ^{2}}e^{-\beta E_{B}\left( \frac{\Delta }{2E_{B}}-1\right)
^{2}} &  & \qquad \text{for }\left( \frac{\Delta }{2E_{B}}-1\right)
^{2}>\left( 1-\frac{\varepsilon }{2E_{B}}\right) ^{2}%
\end{array}%
\right. .  \label{rhodelta}
\end{equation}%
From the barrier distribution (\ref{rhodelta}) we can then compute the
probability 
\begin{equation}
Q\left( \varepsilon \right) =\int_{-\infty }^{\varepsilon }d\Delta \ \rho
\left( \Delta |\,\varepsilon \right)  \label{q}
\end{equation}%
that a particle that has entered the interaction region with energy $%
\varepsilon $ will pass out of it as the particle approaches the interface.
Since the integrand in (\ref{q}) only includes values of $\Delta $ less than
or equal to $\varepsilon ,$ only that part of the distribution (\ref%
{rhodelta}) on the second line of the expression is needed. Using this in (%
\ref{q}) the resulting integral gives for $\varepsilon ,\beta ^{-1}\ll
2E_{B} $%
\begin{align}
Q\left( \varepsilon \right) & =\frac{\Gamma \left( \frac{1}{4},\kappa
^{2}\right) \left( \beta E_{B}\right) ^{1/4}}{\sqrt{4\pi }}\sqrt{1-\frac{%
\varepsilon }{2E_{B}}}\text{exp}\left[ \beta E_{B}\left( 1-\frac{\varepsilon 
}{2E_{B}}\right) ^{2}\right]  \notag \\
& \sim \frac{1}{\sqrt{4\pi \beta E_{B}}}\qquad \qquad \beta E_{B}\gg
1,\;E_{B}\gg \varepsilon  \label{qfinal}
\end{align}%
where in the last line we have again used the asymptotic expansion of the
incomplete gamma function for large arguments.

The last distribution needed to calculate the diffusion constant using
Eqs.\thinspace (\ref{Dgamma}) and (\ref{GAMMA}) is the steady-state
distribution $\rho _{u}^{\text{int}}\left( \varepsilon \right) $ of
untrapped particles inside the interaction region. Starting with the
canonical distribution 
\begin{equation}
\rho ^{\text{int}}\left( \varepsilon _{\text{osc}},\varepsilon \right) =%
\sqrt{\frac{\beta ^{3}}{\pi \varepsilon }}e^{-\beta \varepsilon }\ e^{-\beta
\varepsilon _{\text{osc}}}
\end{equation}%
of the combined particle-oscillator system inside an interaction region, we
exclude that portion associated with trapped states by multiplying by an
appropriate step function that incorporates the self-trapping condition of
Eq. (\ref{SelfTrapped}), i.e.,%
\begin{equation}
\rho _{u}^{\text{int}}\left( \varepsilon _{\text{osc}},\varepsilon \right) =%
\sqrt{\frac{\beta ^{3}}{\pi \varepsilon }}e^{-\beta \varepsilon }\ e^{-\beta
\varepsilon _{\text{osc}}}\theta \left( \varepsilon _{\text{osc}%
}-E_{B}+\varepsilon \left[ 1-\frac{\varepsilon }{4E_{B}}\right] \right) .
\end{equation}%
We now integrate over the oscillator energies to find the associated
particle distribution%
\begin{equation}
f_{u}^{\text{int}}\left( \varepsilon \right) =\int_{0}^{\infty }\rho _{u}^{%
\text{int}}\left( \varepsilon _{\text{osc}},\varepsilon \right) d\varepsilon
_{\text{osc}}=\left\{ 
\begin{array}{ccc}
\sqrt{\frac{\beta }{\pi \varepsilon }}e^{-\beta E_{B}}e^{-\beta \varepsilon
^{2}/4E_{B}} &  & 2E_{B}>\varepsilon >0 \\ 
\sqrt{\frac{\beta }{\pi \varepsilon }}e^{-\beta \varepsilon } &  & 
\varepsilon >2E_{B}%
\end{array}%
\right.  \label{untrappeddistribution}
\end{equation}%
which is normalized to the \emph{fraction} $f_{u}^{\text{int}}$ of untrapped
particles in the interaction region in thermal equilibrium. At very low
temperatures $\beta E_{B}\gg 1$, the fraction of untrapped particles with
energy $\varepsilon >2E_{B}$ becomes negligible and we can approximate the
normalized distribution $\rho _{u}\left( \varepsilon \right) $ of untrapped
particles in the interaction region for all positive $\varepsilon $ by the
form it takes for $\varepsilon <2E_{B},$ i.e.,%
\begin{equation}
\rho _{u}^{\text{int}}\left( \varepsilon \right) =\frac{f_{u}^{\text{int}%
}\left( \varepsilon \right) }{\int_{0}^{\infty }f_{u}^{\text{int}}\left(
\varepsilon \right) d\varepsilon }\sim \frac{\Gamma \left( 3/4\right) }{\pi }%
\left( \frac{\beta }{\varepsilon ^{2}E_{B}}\right) ^{1/4}e^{-\beta
\varepsilon ^{2}/4E_{B}}.  \label{rhoein}
\end{equation}%
Using Eqs. (\ref{gamma}), (\ref{GAMMA}), (\ref{qfinal}), and (\ref{rhoein})
the average escape rate can be written%
\begin{align*}
\Gamma & =\frac{1}{2\sigma }\int_{0}^{\infty }\sqrt{2\varepsilon }\rho _{u}^{%
\text{int}}\left( \varepsilon \right) Q\left( \varepsilon \right) \
d\varepsilon =\frac{\Gamma \left( 3/4\right) }{\sqrt{8\sigma ^{2}\pi ^{3}}%
\beta ^{1/4}E_{B}^{3/4}}\int_{0}^{\infty }\ e^{-\beta \varepsilon
^{2}/4E_{B}}\;d\varepsilon \\
& =\sqrt{\frac{E_{B}}{8\sigma ^{2}\pi ^{2}}}\Gamma \left( 3/4\right) \cdot
\left( \beta E_{B}\right) ^{-3/4}.
\end{align*}%
Using (\ref{Dgamma}), the resulting diffusion constant then takes the form%
\begin{equation}
D=D_{L}^{0}\left( \beta E_{B}\right) ^{-3/4}
\end{equation}%
where%
\begin{equation}
D_{L}^{0}=\frac{a^{2}}{2\sigma }\Gamma \left( 3/4\right) \sqrt{\frac{E_{B}}{%
2\pi ^{2}}}.  \label{DL0}
\end{equation}%
As seen in Fig. \ref{FIG3}, this functional form does an excellent job of
describing the temperature dependence of the diffusion constant for injected
charges at low temperatures for a wide range of model parameters.

\section{Diffusion of Carriers in Thermal Equilibrium}

\label{thermal}

The previous sections focus on calculating the diffusion constant for
particles injected into an untrapped state in a thermalized chain of
oscillators. An ensemble of noninteracting carriers in thermal equilibrium
with the oscillators in the chain will contain both itinerant particles,
which will exchange energy with the oscillator system as they undergo
diffusion along the chain, as well as self-trapped particles, with energies
satisfying (\ref{SelfTrapped}) and which remain bound to the same oscillator
forever. Since the diffusion constant for untrapped particles is identically
zero, the ensemble averaged mean-square displacement of a collection of
particles in thermal equilibrium will be characterized by a diffusion
constant that can be written as the product 
\begin{equation}
D=D_{u}\;f_{u}  \label{Dequilibrium}
\end{equation}%
of the diffusion constant for untrapped particles, i.e., the diffusion
constant evaluated in the last two sections for high and low temperatures,
and the fraction%
\begin{equation}
f_{u}=\int_{0}^{\infty }f_{u}\left( \varepsilon \right) d\varepsilon
\end{equation}%
of untrapped particles in the system. To calculate this quantity we note
that in thermal equilibrium the particle is equally likely to be in any cell
in the system. Focusing on the sub-ensemble in which the particle is in a
particular cell, the probability of a particle to be in the interacting
region or in the noninteracting region is given by an appropriate integral 
\begin{align}
P_{\text{int}}& =Z_{0}^{-1}\int_{L}^{L+2\sigma }dq\,\int_{-\infty }^{\infty
}dp\,\int_{-\infty }^{\infty }d\phi \,\int_{-\infty }^{\infty }d\pi \;\rho _{%
\text{int}}\left( q,p,\phi ,\pi \right)  \label{Pint} \\
P_{\text{non}}& =Z_{0}^{-1}\int_{0}^{L}dq\,\int_{-\infty }^{\infty
}dp\,\int_{-\infty }^{\infty }d\phi \,\int_{-\infty }^{\infty }d\pi \;\rho _{%
\text{non}}\left( q,p,\phi ,\pi \right)  \label{Pnon}
\end{align}%
over the Boltzmann factors%
\begin{align}
\rho _{\text{int}}\left( q,p,\phi ,\pi \right) & =\exp \left( -\beta \left[ 
\frac{p^{2}}{2}+\frac{1}{2}\left( \pi ^{2}+\omega ^{2}\phi ^{2}\right)
+\alpha \phi \right] \right) \\
\rho _{\text{non}}\left( q,p,\phi ,\pi \right) & =\exp \left( -\beta \left[ 
\frac{p^{2}}{2}+\frac{1}{2}\left( \pi ^{2}+\omega ^{2}\phi ^{2}\right) %
\right] \right)
\end{align}%
associated with the Hamiltonian in each region, where 
\begin{align}
Z_{0}& =\int_{L}^{L+2\sigma }dq\,\int_{-\infty }^{\infty }dp\,\int_{-\infty
}^{\infty }d\phi \,\int_{-\infty }^{\infty }d\pi \;\rho _{\text{int}}\left(
q,p,\phi ,\pi \right) \\
& +\int_{0}^{L}dq\,\int_{-\infty }^{\infty }dp\,\int_{-\infty }^{\infty
}d\phi \,\int_{-\infty }^{\infty }d\pi \;\rho _{\text{non}}\left( q,p,\phi
,\pi \right)
\end{align}%
is a normalization factor (or single cell partition function) that makes $P_{%
\text{int}}+P_{\text{non}}=1$. Note that within each region the integrand in
(\ref{Pint}) and (\ref{Pnon}) is independent of $q$, so that the resulting $%
q $-integration just gives a multiplicative factor proportional to the width
of the corresponding region. A straightforward evaluation of the remaining
integrals gives%
\begin{equation}
P_{\text{int}}=\frac{2\sigma }{Le^{-\beta E_{B}}+2\sigma }\qquad \qquad P_{%
\text{non}}=\frac{Le^{-\beta E_{B}}}{Le^{-\beta E_{B}}+2\sigma }.
\label{Pboth}
\end{equation}

\begin{figure}[ptb]
\center \includegraphics[height=2.5in]{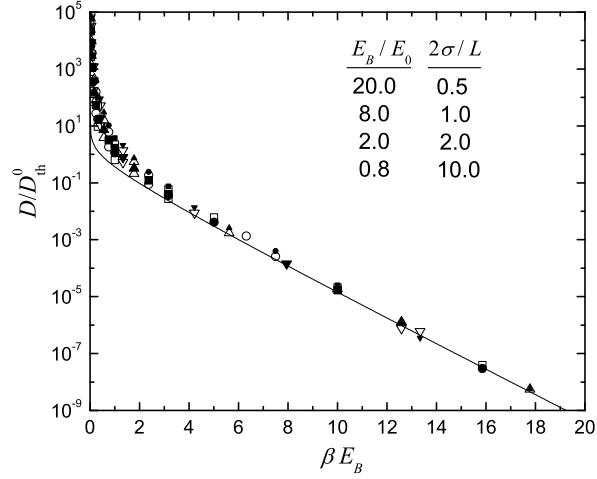}
\caption{Diffusion constant for an ensemble of particles in equilibrium with
the oscillator chain at temperature $T$ as a function of the inverse
temperature $\protect\beta E_{B}$, scaled by the value $D_{th}^{0}$ derived
in the text. Data points include representatives from all sixteen
combinations of the two sets of parameters indicated in the figure. The
solid line is the theoretical expression derived in the text for the
behavior of the diffusion constant at low temperatures.}
\label{FIG5}
\end{figure}

As might be expected, at high temperatures the fraction of particles in each
region is simply proportional to the width of that region. At low
temperatures, by contrast, it becomes exponentially more likely to find the
particle in the interaction region than in the non-interacting region. In
terms of these probabilities, the fraction of untrapped particles in the
ensemble can then be written%
\begin{equation}
f_{u}=P_{\text{int}}f_{u}^{\text{int}}+P_{\text{non}}  \label{fP}
\end{equation}%
where $f_{u}^{\text{int}},$ the fraction of those particles in the
interaction region that are untrapped in equilibrium, may be obtained by
integrating Eq.(\ref{untrappeddistribution}). The result is%
\begin{align}
f_{u}^{\text{int}}& =\int_{0}^{2\beta E_{B}}\sqrt{\frac{\beta }{\pi
\varepsilon }}\exp \left( -\beta \varepsilon \right) \exp \left( -\beta
E_{B}\left( 1-\frac{\varepsilon }{2E_{B}}\right) ^{2}\right) d\varepsilon
+\int_{2\beta E_{B}}^{\infty }\sqrt{\frac{\beta }{\pi \varepsilon }}\exp
\left( -\beta \varepsilon \right) d\varepsilon \\
& =1-\text{erf}\left( \sqrt{2\beta E_{B}}\right) +\left[ \frac{\sqrt{\pi }}{%
\Gamma \left( 3/4\right) }-\frac{\Gamma \left( 1/4,\beta E_{B}\right) }{%
\sqrt{2\pi }}\right] \left( \beta E_{B}\right) ^{1/4}\exp \left( -\beta
E_{B}\right)  \label{fexact}
\end{align}%
which has the following limiting behavior 
\begin{align}
f_{u}^{\text{int}}& \sim \frac{\pi ^{1/2}}{\Gamma \left( 3/4\right) }\left(
\beta E_{B}\right) ^{1/4}\exp \left( -\beta E_{B}\right) \qquad \qquad \text{%
for\ }\beta E_{B}\gg 1  \label{flow} \\
f_{u}^{\text{int}}& \sim 1-\frac{16}{15}\sqrt{\frac{2}{\pi }}\left( \beta
E_{B}\right) ^{3/2}\qquad \qquad \qquad \qquad \;\;\;\;\text{for\ }\beta
E_{B}\ll 1  \label{fhigh}
\end{align}%
at high and low temperatures. Clearly at high temperatures the fraction of
trapped particles is small, and the resulting diffusion constant for an
ensemble of particles in thermal equilibrium will follow the behavior given
in Eq.(\ref{HT}), but with algebraic corrections. At low temperatures,
however, Eqs.(\ref{Pboth}),(\ref{fP}), and (\ref{flow}) show that an
exponentially small fraction 
\begin{equation}
f_{u}\sim \frac{\pi ^{1/2}}{\Gamma \left( 3/4\right) }\left( \beta
E_{B}\right) ^{1/4}\exp \left( -\beta E_{B}\right)
\end{equation}%
of particles are in untrapped states. As a result, the diffusion constant as
given by Eq. (\ref{Dequilibrium}) becomes exponentially suppressed, and will
follow the form 
\begin{equation}
D\sim D_{\text{th}}^{0}\frac{\exp \left( -\beta E_{B}\right) }{\sqrt{\beta
E_{B}}}\qquad \qquad \beta E_{B}\gg 1  \label{Dthermlow}
\end{equation}%
where%
\begin{equation}
D_{\text{th}}^{0}=\frac{a^{2}}{2\sigma }\sqrt{\frac{E_{B}}{2\pi }}.
\label{Dlow0}
\end{equation}%
In Fig. (\ref{FIG5}) we show numerical data for the diffusion constant of an
ensemble of carriers in equilibrium with an oscillator chain as a function
of inverse temperature, plotted on a logarithmic scale to show the
exponential suppression of the diffusion constant as $T\rightarrow 0$. The
solid curve is the low temperature limiting form in Eq. (\ref{Dthermlow}).

\section{Conclusion}

In this paper we have introduced a simple classical version of the Holstein
polaron, namely an otherwise free mobile particle that experiences a linear,
local interaction with vibrational modes distributed throughout the medium
in which it moves. Similar to what has been observed in quantum mechanical
treatments of the problem we find that there are two different types of
state of the combined system: self-trapped and itinerant. Self-trapped
states include particle-oscillator states at negative energy, as well as
some states at positive energies less than the polaron binding energy $E_{B}$%
. As their name suggests, self-trapped states are immobile. Itinerant
polaron states, by contrast, undergo motion that is macroscopically
diffusive. The diffusion constant for such states varies as a power of the
temperature, with two different regimes similar to the adiabatic and
non-adiabatic limits discussed in the polaron literature. At high
temperatures the diffusion constant for untrapped particles varies with
temperature as $\left( \beta E_{B}\right) ^{-5/2}$ and the transport
mechanism is similar to what one thinks of as band transport in a solid,
with long excursions before a velocity reversing scattering from one of the
oscillators in the system. At low temperatures transport occurs via hopping
of the carrier between different cells, with the particle spending a long
time in a given cell before making a transition to a neighboring one. In
this limit the diffusion constant for untrapped particles varies as $\left(
\beta E_{B}\right) ^{-3/4}$.

An ensemble of particles in equilibrium with the chain exhibits a diffusion
constant that is exponentially activated at low temperatures, with an
activation energy equal to the polaron binding energy $E_{B}$. The
functional form, i.e., exponentially activated with an algebraic prefactor,
is reminiscent of the one that has been derived using the small polaron
hopping rate for a particle moving through a tight-binding band of states 
\cite{Holstein}. We anticipate that any mechanism (not included in the
present model) that would allow for a slow equilibration or exchange of
itinerant and self-trapped particle states would not affect the functional
dependence of the diffusion constant with temperature as we have derived in
this paper. The present model we believe is simple enough that many aspects
of the statistical mechanics of the system in equilibrium can be derived
exactly, a circumstance that will facilitate an analysis of transport
properties. In a future paper we consider nonequilibrium aspects of this
model in an attempt to study, from a microscopic point of view, various
aspects of the nonequilibrium statistical mechanics associated with it.

Indeed, there has been continuing interest in the theoretical problem of
deriving macroscopic transport laws (such as Ohm's law) from microscopic,
Hamiltonian dynamics \cite{cels,PhysicaD}. This problem is directly related
to the Hamiltonian description of friction and diffusion, since any such
model must describe the dissipation of the energy of a particle moving
through a medium with which it can exchange energy and momentum, and must
predict the temperature and parameter dependence of transport coefficients,
which will depend on the model considered and on its microscopic dynamics.
The model we have studied here is one of a class of models in which a
particle, moving through a $d$-dimensional space, interacts \emph{locally}
with vibrational \cite{bdb} (or rotational \cite{mex,eck}) degrees of
freedom representing the medium, with the total energy being conserved
during the evolution of the non-linear coupled particle-environment system.
We have investigated the emergence of microscopic chaos arising in the local
dynamics of the current model elsewhere \cite{DPS}. For another such model 
\cite{bdb}, with an infinite number of degrees of freedom at each point in a 
$d$-dimensional space, it was shown rigorously that, at finite total energy,
and when a constant electric field is applied, the particle reaches an
asymptotic speed proportional to the applied field. Positive temperature
states, which would have infinite energy, were not treated in \cite{bdb}.
Their rigorous analysis would be considerably more difficult. The model
studied in the present paper is of the same type as that in \cite{bdb}, but
the infinite-dimensional and continuously distributed harmonic systems of
the previous work have been replaced by isolated, periodically-placed
oscillators. This, as we have seen, allows the system to be studied
numerically and to some extent analytically at positive temperature.

As we have suggested in the introduction, our system can also be thought of
as a $1$D inelastic Lorentz gas (or Pinball Machine). Recall that in the
usual periodic Lorentz gas, circular scatterers are placed periodically on a
plane, with the particle moving freely between them, and scattering
elastically upon contact with the obstacles. For this system, which has been
studied extensively, it has been proven that particle motion in the absence
of an external field is diffusive \cite{bs}. The Lorentz gas, however,
cannot describe the dissipation of energy of the particle into its
environment, since all collisions are elastic. To remedy this situation, the
use of a Gaussian thermostat, artificially keeping the particle kinetic
energy constant during the motion, has been proposed \cite{mh} and proven to
lead to Ohm's law \cite{cels}. Rather than using a thermostat, we have
chosen in the present study to impart internal degrees of freedom to the
obstacles, and to treat explicitly the Hamiltonian dynamics of the coupled
particle-obstacle system, in a spirit similar to \cite{bdb} and the more
complicated two-dimensional models of \cite{mex,eck}. This, as we have
shown, leads to a rich, but straightforwardly characterized temperature and
model parameter dependence of the diffusion constant of the interacting
system.


\begin{thebibliography}{99}
\bibitem{footnote1} This Hamiltonian can be obtained from a more general one
in which the particle and oscillator masses are explicitly included by an
appropriate scale transformation, see, e.g., Ref. \cite{DPS}.
\bibitem{footnote2} The mean-square displacement of many trajectories were
collected into bins of appropriately increasing width to produce the evenly
spaced data on the log-log plots of Fig. \ref{FIG2}a or linear plots of Fig. %
\ref{FIG2}b).
\end{thebibliography}
\end{document}